\documentclass[aps,prb,amsmath,amssymb,amsfonts,preprint,superscriptaddress,citeautoscript,byrevtex]{revtex4-1}
\usepackage{graphicx,color}
\usepackage{epstopdf}
\usepackage{lineno}
\usepackage{soul}
\usepackage{ulem}

\begin{document}
\title[title]{
Magnetic tunnel junctions with a $B2$-ordered CoFeCrAl equiatomic Heusler alloy
}
\author{Tomoki Tsuchiya}
\email{tomoki.tsuchiya.d1@tohoku.ac.jp}
\affiliation{Center for Science and Innovation in Spintronics (CSIS), Core Research Cluster (CRC), Tohoku University, Sendai 980-8577, Japan}
\affiliation{Center for Spintronics Research Network (CSRN), Tohoku University, Sendai 980-8577, Japan}
\author{Tufan Roy}
\affiliation{Research Institute of Electrical Communication (RIEC), Tohoku University, Sendai 980-8579, Japan}
\author{Kelvin Elphick}
\affiliation{Department of Electronics, University of York, York YO10 5DD, England}
\author{Jun Okabayashi}
\affiliation{Research Center for Spectrochemistry, University of Tokyo, Tokyo 113-0033, Japan}
\author{Lakhan Bainsla}
\affiliation{WPI Advanced Institute for Materials Research (AIMR), Tohoku University, Katahira 2-1-1, Sendai 980-8577, Japan}
\author{Tomohiro Ichinose}
\affiliation{WPI Advanced Institute for Materials Research (AIMR), Tohoku University, Katahira 2-1-1, Sendai 980-8577, Japan}
\author{Kazuya Suzuki}
\affiliation{WPI Advanced Institute for Materials Research (AIMR), Tohoku University, Katahira 2-1-1, Sendai 980-8577, Japan}
\affiliation{Center for Spintronics Research Network (CSRN), Tohoku University, Sendai 980-8577, Japan}
\author{Masahito Tsujikawa}
\affiliation{Research Institute of Electrical Communication (RIEC), Tohoku University, Sendai 980-8579, Japan}
\affiliation{Center for Spintronics Research Network (CSRN), Tohoku University, Sendai 980-8577, Japan}
\author{Masafumi Shirai}
\affiliation{Research Institute of Electrical Communication (RIEC), Tohoku University, Sendai 980-8579, Japan}
\affiliation{Center for Science and Innovation in Spintronics (CSIS), Core Research Cluster (CRC), Tohoku University, Sendai 980-8577, Japan}
\affiliation{Center for Spintronics Research Network (CSRN), Tohoku University, Sendai 980-8577, Japan}
\author{Atsufumi Hirohata}
\affiliation{Department of Electronics, University of York, York YO10 5DD, England}
\author{Shigemi Mizukami}
\email{shigemi.mizukami.a7@tohoku.ac.jp}
\affiliation{WPI Advanced Institute for Materials Research (AIMR), Tohoku University, Katahira 2-1-1, Sendai 980-8577, Japan}
\affiliation{Center for Science and Innovation in Spintronics (CSIS), Core Research Cluster (CRC), Tohoku University, Sendai 980-8577, Japan}
\affiliation{Center for Spintronics Research Network (CSRN), Tohoku University, Sendai 980-8577, Japan}
\date{\today}

\begin{abstract}
The equiatomic quaternary Heusler alloy CoFeCrAl is a candidate material for spin-gapless semiconductors (SGSs). However, to date, 
there have been no experimental attempts at fabricating a junction device.
This paper reports a fully epitaxial (001)-oriented MgO barrier magnetic tunnel junction (MTJ) with CoFeCrAl electrodes grown on a Cr buffer.
X-ray and electron diffraction measurements show that the (001) CoFeCrAl electrode films with atomically flat surfaces have a $B2$-ordered phase.
The saturation magnetization is 380 emu/cm$^3$, almost the same as the value given by the Slater--Pauling--like rule,
and the maximum tunnel magnetoresistance ratios at 300 K and 10 K are 87\% and 165\%, respectively.
Cross-sectional electron diffraction analysis shows that the MTJs have MgO interfaces with fewer dislocations.
The temperature- and bias-voltage-dependence of the transport measurements indicates magnon-induced inelastic electron tunneling 
overlapping with the coherent electron tunneling.
X-ray magnetic circular dichroism (XMCD) measurements show a ferromagnetic arrangement of the Co and Fe magnetic moments of $B2$-ordered CoFeCrAl, in contrast to the ferrimagnetic arrangement predicted for the $Y$-ordered state possessing SGS characteristics.
Ab-initio calculations taking account of the Cr-Fe swap disorder qualitatively explain the XMCD results.
Finally, the effect of the Cr-Fe swap disorder on the ability for electronic states to allow coherent electron tunneling is discussed.
\end{abstract}

\maketitle
\section{INTRODUCTION}
A spin-gapless semiconductor (SGS) is a material in which the Fermi level is located at a zero-energy gap state for a majority spin band and at an energy gap for a minority spin band.\cite{Wang2008,Wang2010,Wang2017}
SGSs belong to the class of half-metals that have fully spin-polarized carriers at the Fermi level, 
so they exhibit a huge magnetoresistance (MR) and low spin relaxation (the so-called Gilbert damping).
These physical properties are ideally suited to solid-state spintronic devices,
and are commonly observed in half-metals.\cite{Sakuraba2006,Iwase2009,Mizukami2001,Kubota2009, Bainsla2018a,Kudrnovsky2018,Bainsla2018b} 
In addition to such physical properties, SGSs could be used to realize devices with new functionalities, 
such as reconfigurable magnetic tunnel diodes and transistors,\cite{Şaşioglu2016}
which use their gapless electronic characteristics.
Therefore, it is of fundamental and technological importance to investigate such advanced spintronic materials.

Many candidate materials for SGSs have been proposed. 
One candidate is an equiatomic quaternary Heusler alloy (EQHA) with a chemical formula of XX'YZ,\cite{Xu2013,Ozdogan2013}
where X, X', and Y denote transition metal elements and Z represents a main group element.
The crystal structure of EQHAs is a cubic LiMgPdSn or $Y$-type, 
as shown in Fig. 1(a). 
Because there are various possible arrangements of the elements, 
EQHAs exhibit several chemically disordered structures, {\it e.g.}, the $XA$-type, which belongs to the same space group as the $Y$-type [Fig. 1(b)] and 
the $L2_1$-, $B2$-, and $A2$-types, which have different space groups [Figs. 1(b)--1(f)].
In recent years, intensive theoretical and experimental studies have considered various EQHAs.\cite{Bainsla2016}
Their results indicate that, to realize SGSs, it is of vital importance to characterize the chemical orderings of EQHAs and understand their effect on both the gapless state and half-metallic gap.
\begin{figure*}
\begin{center}
\includegraphics[width=12.0cm,keepaspectratio,clip]{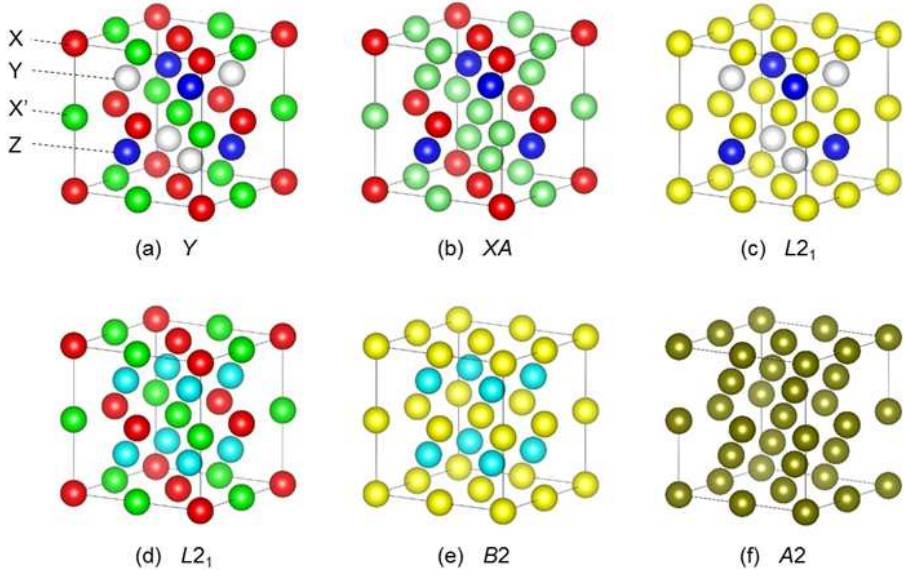}
\caption{
Schematic illustrations of the cubic crystal structure with various chemical orderings for EQHAs, denoted by the chemical formula XX'YZ.
(a) $Y$ (space group $F\bar{4}3m$, No. 216), in which X, X', Y, and Z elements correctly occupy each Wyckoff position of $4a$, $4b$, $4c$, and $4d$, respectively. (b) $XA$ (space group $F\bar{4}3m$, No. 216), in which X and Y are randomly swapped. (c) [(d)] $L2_1$ (space group $F\bar{m}3m$, No. 225), in which X and X' (Y and Z) are randomly swapped. (e) $B2$ (space group $P\bar{m}3m$, No. 221), in which X and X' and also Y and Z are randomly swapped. (e) $A2$ (space group $I\bar{m}3m$, No. 229), in which all elements are randomly swapped. 
}
\end{center}
\end{figure*}

Hereafter, we focus on CoFeCrAl as a typical candidate EQHA for SGSs.
Xu et al. were the first to theoretically suggest that several EQHAs, including CoFeCrAl, would have the abovementioned electronic structure of SGSs.\cite{Xu2013} 
Subsequently, Ozdogan et al. theoretically studied the electronic structure of 60 EQHAs and confirmed that CoFeCrAl becomes an SGS.\cite{Ozdogan2013}
Many experimental and theoretical studies on CoFeCrAl have since been reported.\cite{Luo2009,Gao2013,Nehra2013,Iyigor2014,Alhaj2014,Bainsla2015a,Kharel2015,Jin2016,Choudhary2016,Jin2017, Bhat2017,Bhat2018} 
Luo et al. conducted experiments on bulk samples of CoFeCrAl with the $B2$ chemical ordering,\cite{Luo2009}
reporting a lattice parameter of 0.5760 nm and Curie temperature $T_{c}$ of 460 K.
The saturation magnetic moment $m$ was 2.070 $\mu_B$/f.u. at 5 K, and they suggested that the total spin magnetic moment $m_{\rm s}$ obeys the Slater--Pauling--like rule of half-metallic Heusler alloys.\cite{Luo2009} 
Nehra et al. reported similar results.\cite{Nehra2013}
Subsequently, Bainsla et al. obtained $B2$-ordered CoFeCrAl bulk samples in which $m = 2$ $\mu_B$/f.u.,
and their samples exhibited a metallic temperature-dependence in resistivity and a maximum transport spin polarization $P_T$ of 64\%, as evaluated by a point-contact Andreev reflection (PCAR) technique.\cite{Bainsla2015a} 
In contrast, Kharel et al. reported non-metallic temperature-dependence in the resistivity for CoFeCrAl bulk ribbon samples prepared by a melt spinning technique.\cite{Kharel2015}
Their samples exhibited very weak superlattice peaks stemming from the $L2_1$ chemical ordering,
indicating that the chemical ordering is better than the $B2$ ordering.\cite{Kharel2015}
They reported $m$ values of 1.9 and 2.1 $\mu_B$/f.u. and $T_{c}$ values of 456 and 540 K, respectively, for samples annealed under different conditions, and discussed these results in terms of the zero-gap electronic states smeared by the chemical disorder.\cite{Kharel2015}
Later, the same group studied CoFeCrAl epitaxial thin films grown on MgO substrates using a sputtering deposition technique.\cite{Jin2016,Jin2017}
They reported that the films exhibited the $L2_1$ chemical order, and measured $m$ = 2.0 $\mu_B$/f.u., $T_{c}$ = 390 K, a semimetal-like carrier number density of 1.2$\times$10$^{18}$ cm$^{-3}$, and $P_T$ = 68\%.\cite{Jin2016}
The observed results were discussed in terms of the SGS characteristics.\cite{Jin2016}
To date, there have been no experimental studies on magnetic tunnel junctions (MTJs),
which are important because a huge tunnel magnetoresistance (TMR) effect is expected from the high spin polarization of CoFeCrAl.

In this paper, we describe the spin-dependent transport properties of fully epitaxial MTJs with CoFeCrAl epitaxial electrode films.
Previously, we reported the structural and magnetic properties of epitaxial films of CoFeMnSi,
which is another EQHA that is an SGS candidate.
The films grown on a Cr buffer had a $B2$ as well as partial $L2_1$ orderings,\cite{Bainsla2017}
and their MTJs exhibited TMR ratios of more than 500\% at 10 K, 
suggesting half-metallic electronic characteristics.\cite{Bainsla2018a}
Different from CoFeMnSi, only $B2$-ordered CoFeCrAl films were obtained in this study, despite the similar fabrication conditions and vacuum deposition apparatus.
The observed TMR ratios for MTJs in the CoFeCrAl electrode films were 87\% at 300 K and 165\% at 10 K,
even though the abovementioned $P_T$ values for CoFeCrAl are not much different from that of CoFeMnSi ($P_T$ = 64\%).\cite{Bainsla2015b}
%The temperature dependence of this TMR ratio was much smaller than that for the MTJs with CoFeMnSi,
%though the above-mentioned values of $T_{c}$ for CoFeCrAl is about 100-200 K lower than that for CoFeMnSi, 623 K.\cite{Alijani2011}
The underlying physics and chemistry are discussed based on both microscopic characterizations of the interface structure and elemental magnetism and ab-initio calculations that take account of various chemical disorders.

\section{EXPERIMENTAL AND THEORETICAL CALCULATION PROCEDURES}
All samples were deposited on MgO(100) single-crystal substrates using a magnetron sputtering technique. 
The base pressure of the deposition chamber was 2$\times$10$^{-7}$ Pa. 
The MTJ staking structure was substrate/Cr(40)/CoFeCrAl(30)/Mg(0.4)/MgO(2)/CoFe(5)/ IrMn(10)/Ta(3)/Ru(5) (thickness is in nanometers). 
Before the deposition, the surfaces of the substrates were cleaned by flushing at 700$^\circ$C in the chamber. 
All layers were deposited at room temperature (RT). 
The Cr buffer layer was annealed in situ at 700$^\circ$C for 1 h to obtain a flat surface with (001) orientation.\cite{Bainsla2017} 
The CoFeCrAl layer was deposited on the substrate using an alloy target, 
with the film composition of Co$_{25.5}$Fe$_{23.1}$Cr$_{28.1}$Al$_{23.3}$ (at.\%) determined 
using an inductively coupled plasma mass spectrometer. After the deposition of the CoFeCrAl layer, 
in situ annealing was performed at temperatures $T_{\rm anneal}$ of 400--800$^\circ$C. 
We also prepared samples of substrate/Cr(40)/CoFeCrAl(30)/Ta(3) for structural and magnetization measurements and samples of substrate/Cr(40)/CoFeCrAl(30)/Mg(0.4)/MgO(2) for x-ray magnetic circular dichroism (XMCD) studies. 

The microfabrication of MTJs with junction areas ranging from 10$\times$10 to 30$\times$30 $\mu$m$^2$ was performed using standard ultraviolet photo-lithography and Ar ion milling. 
Following the microfabrication, ex situ annealing was performed with a vacuum furnace at temperatures $T_{\rm MTJ}$ of 250--500$^\circ$C under an in-plane magnetic field of 5 kOe.
\begin{figure}
\begin{center}
\includegraphics[width=7cm,keepaspectratio,clip]{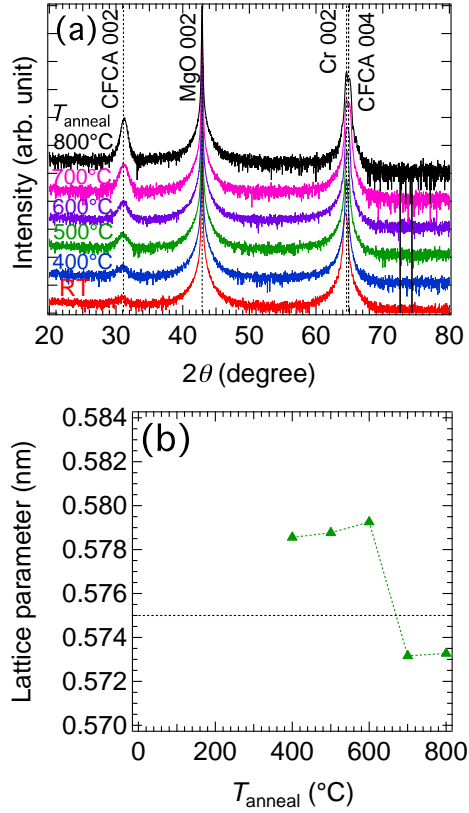}
\caption{
Structural properties of 30-nm-thick CoFeCrAl films. (a) Out-of-plane XRD patterns. (b) Lattice parameters for perpendicular-to-plane ($c$-axis) directions. 
Thin dashed line denotes the bulk value.\cite{Bainsla2015a}
}
\end{center}
\end{figure}

The crystal structures of the samples were determined by x-ray diffraction (XRD) using Cu $K_\alpha$ radiation. 
The surface morphology and roughness were probed by atomic force microscopy (AFM). 
Microstructure analysis was conducted by transmission electron microscopy (TEM). 
Cross-sectional TEM images were used to analyze the crystalline structures of both samples. 
TEM specimens were prepared by hand polishing until the sample thickness became approximately 10 $\mu$m. 
The specimens were then thinned using the Precision Ion Polishing System (PIPS) 
until they became electron-transparent, typically $\sim$100 nm. 
During the ion beam thinning process, the ion gun voltage was operated at 3--5 keV with an incident beam angle of 4--6$^\circ$ depending on the specimen thickness. 
Magnetization measurements were performed using a vibrating sample magnetometer.
Out-of-plane magnetization was measured by the polar magneto-optical Kerr effect (MOKE) with a laser wavelength of about 400 nm.
XMCD measurements were performed at BL-7A in the Photon Factory (KEK). 
Photon helicity was fixed, and a magnetic field switching between $\pm$10 kOe was applied along the incident polarized soft x-ray. 
The extent of circular polarization was evaluated to be 85\%. 
The total-electron-yield mode was adopted. 
The measurements were carried out in a grazing incidence setup with respect to the sample surface normal
in order to detect the in-plane spin and orbital magnetic moments. 
All of the abovementioned measurements were performed at RT. 

The transport properties of the MTJs were investigated using a four-probe method and a prober system 
with a maximum applied field of 1 kOe at RT and a physical property measurement system (PPMS)
at temperatures $T$ ranging from 10--300 K with an applied magnetic field of up to 1 kOe. 
The MTJs with varying junction areas were measured; however, all the data presented here were obtained 
with a junction area of 10$\times$10 $\mu$m$^2$. 

Ab initio calculations were carried out using the full potential spin-polarized-relativistic Korringa--Kohn--Rostoker (FP-SPRKKR) method, 
as implemented in the SPR-KKR program package.\cite{7} 
The effect of substitutional disorder has been considered by coherent potential approximation. 
For the exchange correlation functional, the generalized gradient approximation, as parameterized by Perdew, Burke, and Ernzerhof (PBE), was used.\cite{8} 
An angular expansion of up to $l_{\rm max} = 3$ has been considered for each atom. 
We employed Lloyd’s formula to determine the Fermi energy.\cite{9,10}
We have used 917 irreducible $k$-points for the Brillouin zone integrations.
\begin{figure*}
\begin{center}
\includegraphics[width=14cm,keepaspectratio,clip]{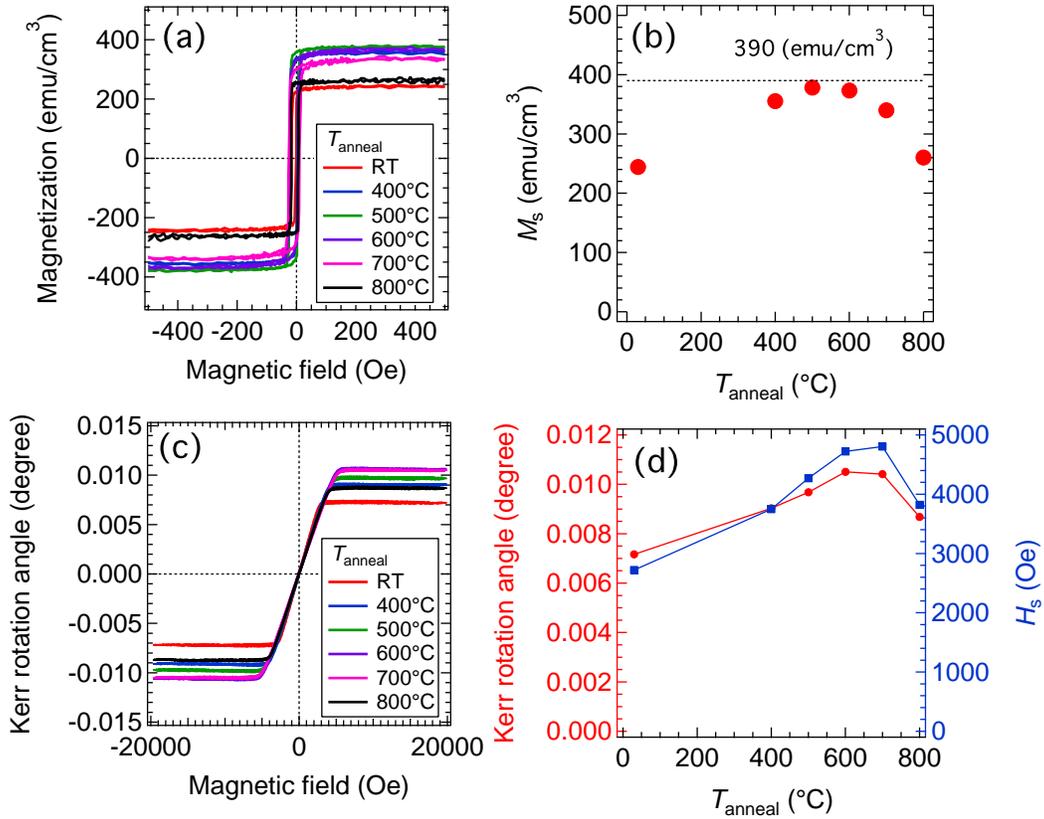}
\caption{
(a) In-plane magnetization hysteresis loops for samples with different annealing temperatures. (b) Saturation magnetization $M_{\rm s}$ as a function of $T_{\rm anneal}$ for CoFeCrAl layer. (c) Polar MOKE curves for the same samples under out-of-plane magnetic field. (d) Saturation Kerr rotation angles and out-of-plane saturation field $H_s$ as a function of $T_{\rm anneal}$ for CoFeCrAl layer. 
}
\end{center}
\end{figure*}

\section{RESULTS AND DISCUSSION}
\subsection{Structure and magnetism for the CoFeCrAl epitaxial films grown on Cr buffer}
\begin{figure*}
\begin{center}
\includegraphics[width=14cm,keepaspectratio,clip]{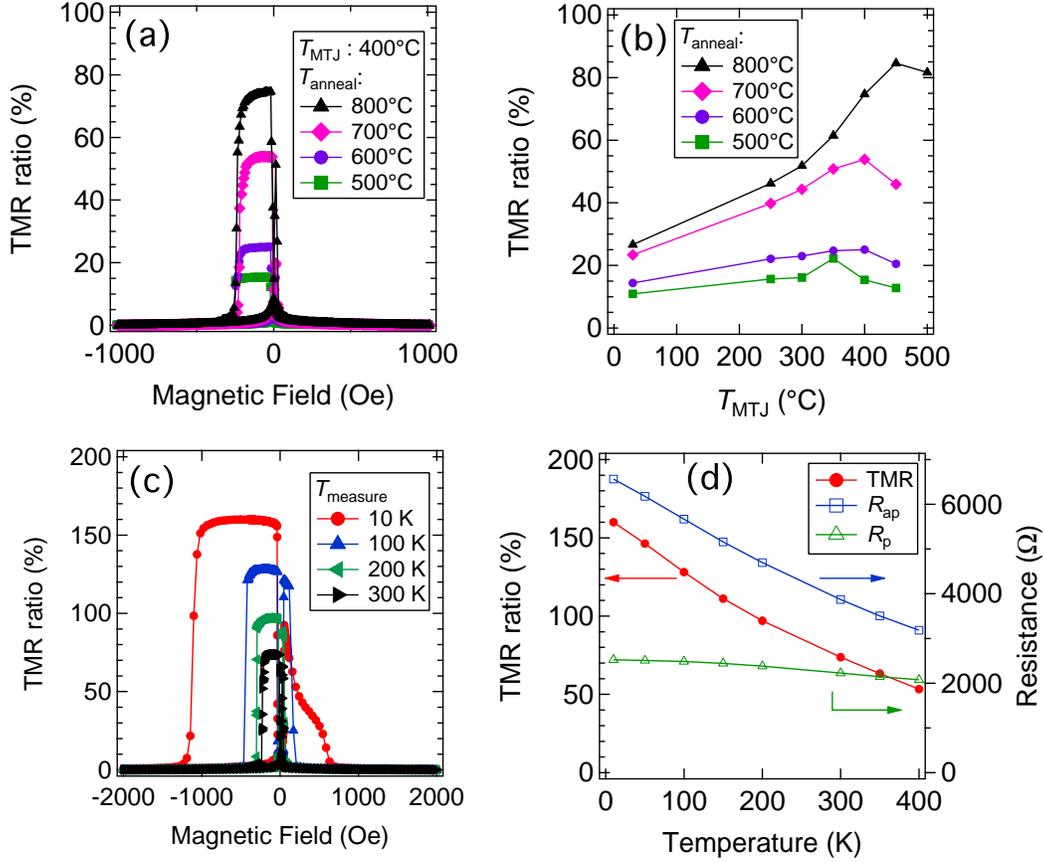}
\caption{
(a) MR curves measured at RT with various $T_{\rm anneal}$. (b) TMR ratio as a function of $T_{\rm MTJ}$. (c) MR curves measured at various temperatures $T$ for the MTJ with $T_{\rm anneal}$ of 800$^\circ$C. (d) TMR ratio as a function of the measurement temperature $T$ for the MTJ with $T_{\rm anneal}$ of 800$^\circ$C.
}
\end{center}
\end{figure*}
Out-of-plane XRD patterns of the CoFeCrAl films are shown in Fig. 2(a). 
All samples exhibit a 002 peak from the Cr buffer layer and a 002 superlattice diffraction peak from the CoFeCrAl. 
No (111) superlattice peaks were observed in any of the samples in the measurement with $\chi =54.74^\circ$ (not shown here). 
These results suggest that all samples have the $B2$ phase, and no $Y$ or $L2_1$ ordered phases. 
The lattice parameter along the $c$-axis is plotted as a function of $T_{\rm anneal}$ in Fig. 2(b).
The lattice parameter of the $c$-axis was calculated from the 002 peak. 
The lattice parameter for a bulk sample is provided for comparison.\cite{Bainsla2015a}
The lattice parameters of the CoFeCrAl films are larger than the bulk value for $T_{\rm anneal}$ below 600$^\circ$C, 
and slightly smaller and nearly constant for $T_{\rm anneal}$ above 600$^\circ$C.
The $c$ lattice parameters of the CoFeCrAl films for $T_{\rm anneal}$ values of 700 and 800$^\circ$C are approximately 
the same at $\sim$0.5732 nm.
The order parameters could not be calculated because of the overlap between the CoFeCrAl 004 peak and the Cr 002 peak.
However, the increase in intensity for the superlattice diffraction peak at higher $T_{\rm anneal}$ suggests an increase in the degree of order. 

The surface morphology and average roughness $R_a$ of CoFeCrAl were also measured by AFM. 
Atomically flat surfaces with $R_a$ less than 0.20 nm were observed in all samples.

In-plane magnetization curves and the saturation magnetization $M_{\rm s}$ as a function of $T_{\rm anneal}$ are shown in Figs. 3(a) and 3(b), respectively. 
All samples exhibit in-plane magnetic anisotropy, as seen in Fig. 3(a).
The small magnetization at lower $T_{\rm anneal}$ may be caused by an unidentified phase or a disordered phase.
$M_{\rm s}$ increases with rising $T_{\rm anneal}$, probably due to the improvement in the degree of order.
$M_{\rm s}$ then decreases when $T_{\rm anneal}$ is above 600$^\circ$C.
To understand this reduction in $M_{\rm s}$, we also measured MOKE under an out-of-plane magnetic field, as shown in Fig. 3(c).
The MOKE curves show linear behavior around the zero field and a saturation at $\pm$3--5 kOe that depends on $T_{\rm anneal}$.
As seen in Fig. 3(d), the saturation value of the Kerr rotation angle and the saturation field $H_s$ for these samples increase and attain maxima at $T_{\rm anneal}$ = 600--700$^\circ$C.
The Kerr rotation angle is approximately proportional to $M_{\rm s}$ and the light skin length is typically 10--20 nm at 400 nm for the transition metals, so that $M_{\rm s}$ within the light skin depth is almost the same for $T_{\rm anneal} = 600-700^\circ$C.
For $T_{\rm anneal} = 600-800^\circ$C, the interdiffusion of the Cr buffer and CoFeCrAl layers proceeds gradually with increasing $T_{\rm anneal}$, and then the magnetic dead layer of CoFeCrAl at the bottom interface becomes thicker.
This may cause the apparent reduction in $M_{\rm s}$ at $T_{\rm anneal} = 600-800^\circ$C observed in Fig. 3(b).
When no other magnetic anisotropies exist, $H_s$ is determined by the shape anisotropy according to $H_s = - 4 \pi M_{\rm s}$.
The value of $H_s$ for $T_{\rm anneal} = 600-700^\circ$C is $\sim$4.8 kOe [Fig. 3(d)], 
from which $M_{\rm s}$ can be evaluated as $\sim$382 emu/cm$^3$.
This is quite close to the maximum $M_{\rm s}$ of 380 emu/cm$^3$ for the samples 
annealed at 500 and 600$^\circ$C, as seen in Fig. 3(b).
Therefore, the $M_{\rm s}$ value near the film surface for $T_{\rm anneal} \sim 500-700^\circ$C would be similar to 380 emu/cm$^3$,
though it is slightly smaller for $T_{\rm anneal} = 800^\circ$C.
The magnetic moment $m$ evaluated from this $M_{\rm s}$ value is $\sim$1.9 $\mu_B$/f.u., 
which is comparable to the value calculated from the Slater--Pauling-like rule, 2.0 $\mu_B$/f.u. at the ground state,
and is consistent with previous reports.\cite{Luo2009,Nehra2013,Bainsla2015a,Kharel2015,Jin2016}

\subsection{Spin-dependent transport in MTJs with the CoFeCrAl electrode and its interface structures}
The MR curves measured at RT for MTJs with $T_{\rm MTJ}$ = 400$^\circ$C and various $T_{\rm anneal}$ are shown in Fig. 4(a). 
The TMR ratios as a function of $T_{\rm anneal}$ are shown in Fig. 4(b). 
The bias voltage $V$ was $\sim$10 mV for this measurement.
The TMR effect was observed in all samples, and the TMR ratio first increases with increasing $T_{\rm MTJ}$, 
then decreases at certain values of $T_{\rm MTJ}$.
The $T_{\rm MTJ}$ values at which the TMR ratio attains a maximum tend to increase slightly with increasing $T_{\rm anneal}$ [Fig. 4(b)].
Further, the TMR ratios change significantly among the MTJs with different $T_{\rm anneal}$ values of $500-800^\circ$C, as clearly seen in Fig. 4(a).
In this study, the highest TMR ratio observed at 300 K was 87\% for the MTJ with $T_{\rm anneal}$ = 800$^\circ$C 
and $T_{\rm MTJ}$ = 450$^\circ$C.

To clarify the transport mechanism, the $T$-dependence of the TMR effect measured at $V \sim 10$ mV
was investigated for the MTJ with $T_{\rm anneal}$ = 800$^\circ$C and $T_{\rm MTJ}$ = 400$^\circ$C.
The MR curves measured at various temperatures and TMR ratios of CoFeCrAl for $T_{\rm anneal}$ = 800$^\circ$C
are presented as a function of $T$ in Figs. 4(c) and 4(d), respectively.
The MR curves show well-defined parallel (P) and antiparallel (AP) states at all temperatures.
The TMR ratio increases almost linearly with decreasing $T$, as seen in Fig. 4(d),
reaching $\sim$160\% at $T = 10$ K, which is almost twice the value of $\sim$75\% observed at $T = 300$ K.
Furthermore, the junction resistance in the AP state $R_{\rm AP}$ (in the P state $R_{\rm P}$) increases strongly (weakly) with decreasing $T$.
The tendency of $T$-dependence in Fig. 4(d) can be explained by the inelastic electron tunneling due to the magnon,
because our data are similar to those for some CoFeB/MgO/CoFeB and CoFe/MgO/CoFe MTJs
discussed by Drewello et al. in terms of the magnon effect.\cite{Drewello2008}
\begin{figure}
\begin{center}
\includegraphics[width=7.5cm,keepaspectratio,clip]{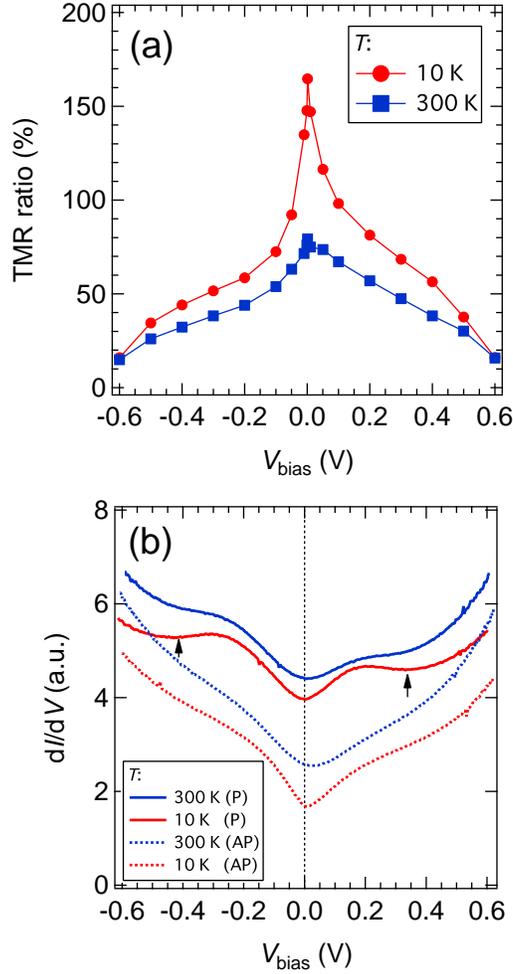}
\caption{
Bias voltage dependence of the spin-dependent transport property measured at 10 and 300 K for the MTJ 
with $T_{\rm anneal}$ of 800$^\circ$C and $T_{\rm MTJ}$ of 400$^\circ$C. (a) TMR ratio, (b) junction resistance at 300 K and 10 K. P and AP denote the parallel and antiparallel states of the magnetizations of CoFe and CoFeCrAl, respectively.
}
\end{center}
\end{figure}

Figure 5(a) shows the TMR ratio as a function of the junction bias $V$ measured at 10 and 300 K for the MTJs 
in the same sample device as in Figs. 4(c) and 4(d).
In addition to the gradual and asymmetric variations with respect to the polarity of $V$,
the TMR ratio exhibits very rapid changes within about $\pm$ 0.1 V at $T = 10$ K.
Figure 5(b) displays the differential conductance data $dI/dV$ vs $V$ measured at 300 and 10 K
for the corresponding MTJs.
The conductance dips near $V=0$ are clearly visible in both the P and AP states for both values of $T$,
and are correlated with the abovementioned large change in the TMR ratio near $V=0$ V.
These zero-bias anomalies are also explained by the inelastic electron tunneling due to the magnon, as mentioned above,
that were observed in the $dI/dV$ data for some CoFeB/MTJ/CoFeB MTJs.\cite{Drewello2008}

As well as the zero-bias dip, we also observed different structures in the $dI/dV$ data of the P state at $V = \pm 0.3$ V, 
as indicated with arrows in Fig. 5(b).
Their positions and shapes are similar to those for the structure observed in the $dI/dV$ data of the P state in CoFe(B)/MgO/CoFe(B) MTJs with Co-rich compositions.\cite{Bonell2012}
In the CoFe(B)/MgO/CoFe(B) MTJs, these structures were considered to result from the coherent tunneling process via spin-polarized $\Delta_1$ band for the tunneling electron wave vector parallel to the [001] direction of CoFe.\cite{Bonell2012}

\begin{figure}
\begin{center}
\includegraphics[width=5.5cm,keepaspectratio,clip]{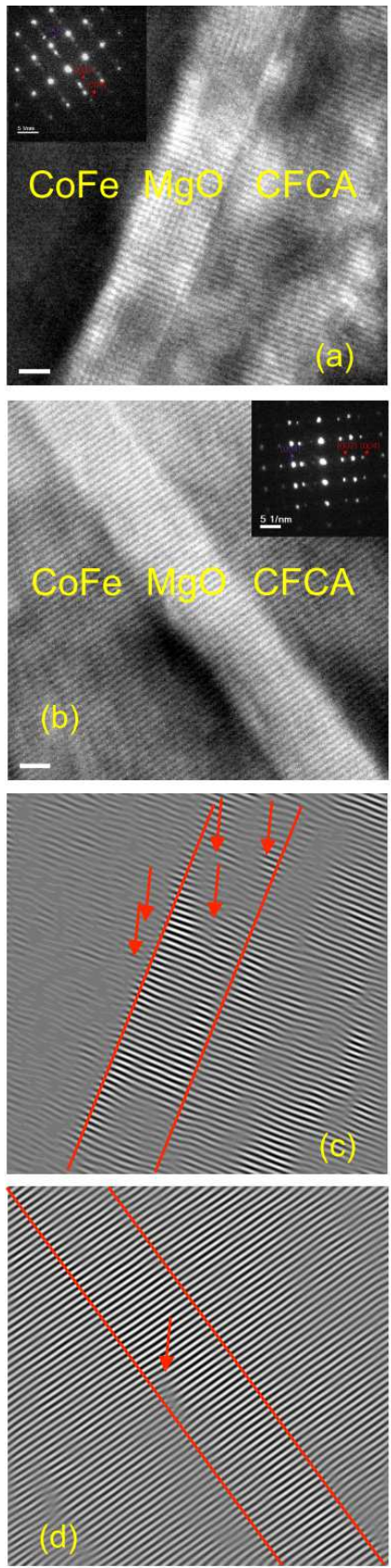}
\caption{
Cross-sectional TEM images of the samples annealed at (a) 500$^\circ$C and (b) 800$^\circ$C. The corresponding selected area beam diffraction patterns are shown as insets. The corresponding lattice planes are also shown for samples annealed at (c) 500$^\circ$C and (d) 800$^\circ$C.
}
\end{center}
\end{figure}
The coherent electron tunneling takes place along the coherent lattices at the heterointerfaces of electrode/barrier/electrode. Thus,
we investigate the nanostructure of the interface for the CoFeCrAl/MgO/CoFe MTJs
via the high-resolution cross-sectional TEM measurements for the two representative samples.
Figures 6(a) and 6(b) show cross-sectional TEM images of the samples with CoFeCrAl electrodes annealed 
at $T_{\rm anneal}=$ 500$^\circ$C and 800$^\circ$C, respectively. 
Based on the measurement of lattice-fringe spacing from the high-resolution TEM image,
the lattice constant of the CoFeCrAl is approximately 0.585 nm. 
Nano-beam diffraction (NBD) patterns were taken to identify the crystalline structures [insets in Figs. 6(a) and 6(b)]. 
Diffraction spots from both CoFeCrAl and MgO layers can be observed in both specimens. 
Strong diffraction spots from the CoFeCrAl (004) plane are detected, which agrees with the XRD data shown above. 
This structural analysis confirms that CoFeCrAl exhibits predominant $B2$ ordering rather than $L2_1$ ordering.
Figures 6(c) and 6(d) show the corresponding crystalline lattice planes between the CoFeCrAl and MgO layers,
with approximately 15 nm across the plane. The images were obtained by selectively displaying crystalline planes across grain boundaries. 
The dislocation of the lattice plane can be identified clearly from the images, as indicated by arrows. 
When a single fringe is split into two, it indicates the presence of lattice dislocations. 
As shown in Fig. 6(c), there are multiple dislocations at the bottom and top CoFeCrAl/MgO/CoFe interface, as well as within the MgO barrier, for the sample with CoFeCrAl annealed at $T_{\rm anneal}=$ 500$^\circ$C.
Note that the top interface has more dislocations than the bottom one. 
In contrast, only one dislocation can be observed in the sample with CoFeCrAl annealed at $T_{\rm anneal}=$ 800$^\circ$C, 
as shown in Fig. 6(d). 
These results confirm that higher values of $T_{\rm anneal}$ reduce the number of dislocations at the CoFeCrAl/MgO interfaces and within the MgO barrier,
and also suggest that the coherent electron tunneling is possible from a structural point of view. 
This is also supportive in explaining the difference in the TMR ratio (by almost a factor of four) between the two samples in terms of the structural imperfections.

\subsection{Microscopic identifications}
To consider the physics underlying the abovementioned transport properties,
it is crucial to experimentally identify the electronic state of the CoFeCrAl electrode near the interface.
Hence, we investigated the elemental magnetic moments using XMCD measurements.
Note that the XMCD measurements typically probe elements within a few nanometers in depth. Hence,
we were able to obtain an insight into the electronic state near the interface of MgO and CoFeCrAl via the XMCD results 
with the aid of ab-initio calculations that take account of possible chemical disorders.

Figures 7(a) and 7(b) show the x-ray absorption spectra (XAS) and XMCD spectra, respectively,
of Cr, Fe, and Co $L_{2,3}$ edges with different photon helicity for the sample annealed at $T_{\rm anneal}=$ 700$^\circ$C. 
Clear metallic peaks can be observed, confirming that there is no mixing of oxygen atoms. 
Shoulder structures appear in the higher-photon-energy region of the Co $L_{2,3}$ XAS peaks. 
These are considered to originate from the Co-Co bonding states in Heusler alloy structures.\cite{Galanakis2002}
No finite XMCD signals can be observed at the Cr $L$-edges [Fig. 7(b)]. 
The XAS and XMCD spectra for the sample without annealing were also measured as a reference (not shown here) 
and were similar to the data in Fig. 7,
except for less pronounced shoulder structures for Co $L_{2,3}$ XAS peaks and less magnetic contrast in XMCD.
This change with the annealing temperature is consistent with the view that the degree of chemical order increases with annealing,
as discussed based on the XRD results.

The spin and orbital magnetic moments were estimated by applying the magneto-optical sum rules.
The magnetic moments given by summing both spin and orbital components of each element
are estimated to be 1.14 and 0.52 $\mu_B$/atom for Fe and Co, respectively, for the $T_{\rm anneal}=$ 700$^\circ$C CoFeCrAl film.
The total magnetic moment $m$ is 1.66 $\mu_B$/f.u., which is similar to the experimental value of $\sim$1.9 $\mu_B$/f.u. stated earlier and the theoretical value of 2.0 $\mu_B$/f.u. for $Y$-ordered CoFeCrAl.
Interestingly, the XMCD results confirmed that the net magnetic moment of Co seems to be ferromagnetically coupled to that of Fe for the samples in this study.
This is dissimilar to the antiferromagnetic arrangement between them that has previously been predicted for the $Y$-ordered case.\cite{Choudhary2016}
This finding is confirmed by the element-specific magnetic hysteresis for Fe and Co shown in Figs. 7(c) and 7(d), respectively.

\begin{figure}
\begin{center}
\includegraphics[width=8cm,keepaspectratio,clip]{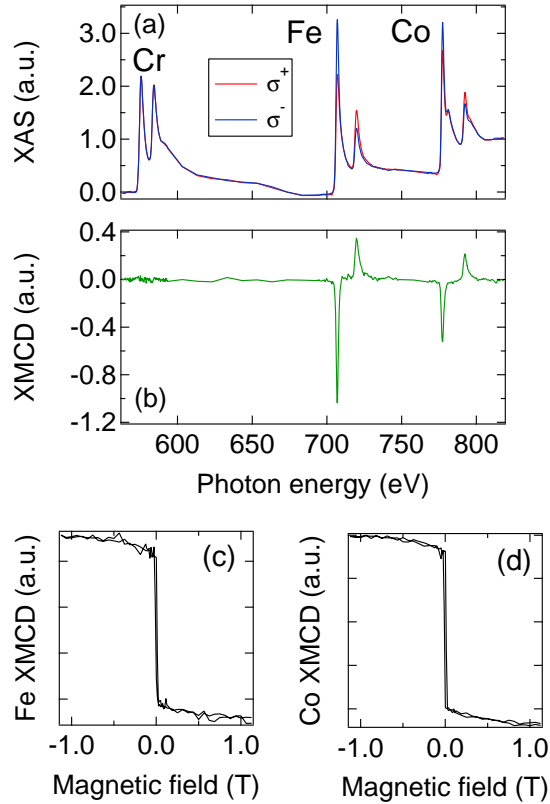}
\caption{
(a) XAS with different magnetic fields for the CFCA sample annealed at 700$^\circ$C
measured at the Cr, Fe, and Co $L_{3,2}$ edges. 
(b) XMCD spectra for the corresponding elements are also shown. 
The in-plane hysteresis curves of XMCD taken at the (c) Fe and 
(d) Co $L_3$-edge.
}
\end{center}
\end{figure}
\begin{figure}
\begin{center}
\includegraphics[width=7cm,keepaspectratio,clip]{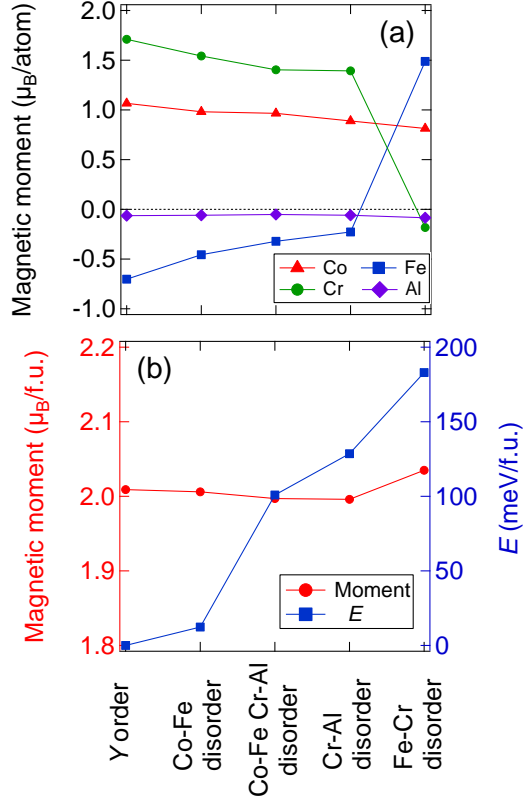}
\caption{
(a) Calculated element-specific magnetic moments and (b) calculated total magnetic moments and the formation energy of CoFeCrAl with various ordering states: (i) full ordering $Y$, (ii) full random swapping of Co and Fe, (iii) full random swapping of Cr and Al, 
(iv) full random swapping of Co and Fe as well as that of Cr and Al, and (v) full random swapping of Fe and Cr.
}
\end{center}
\end{figure}
\begin{figure}
\begin{center}
\includegraphics[width=7cm,keepaspectratio,clip]{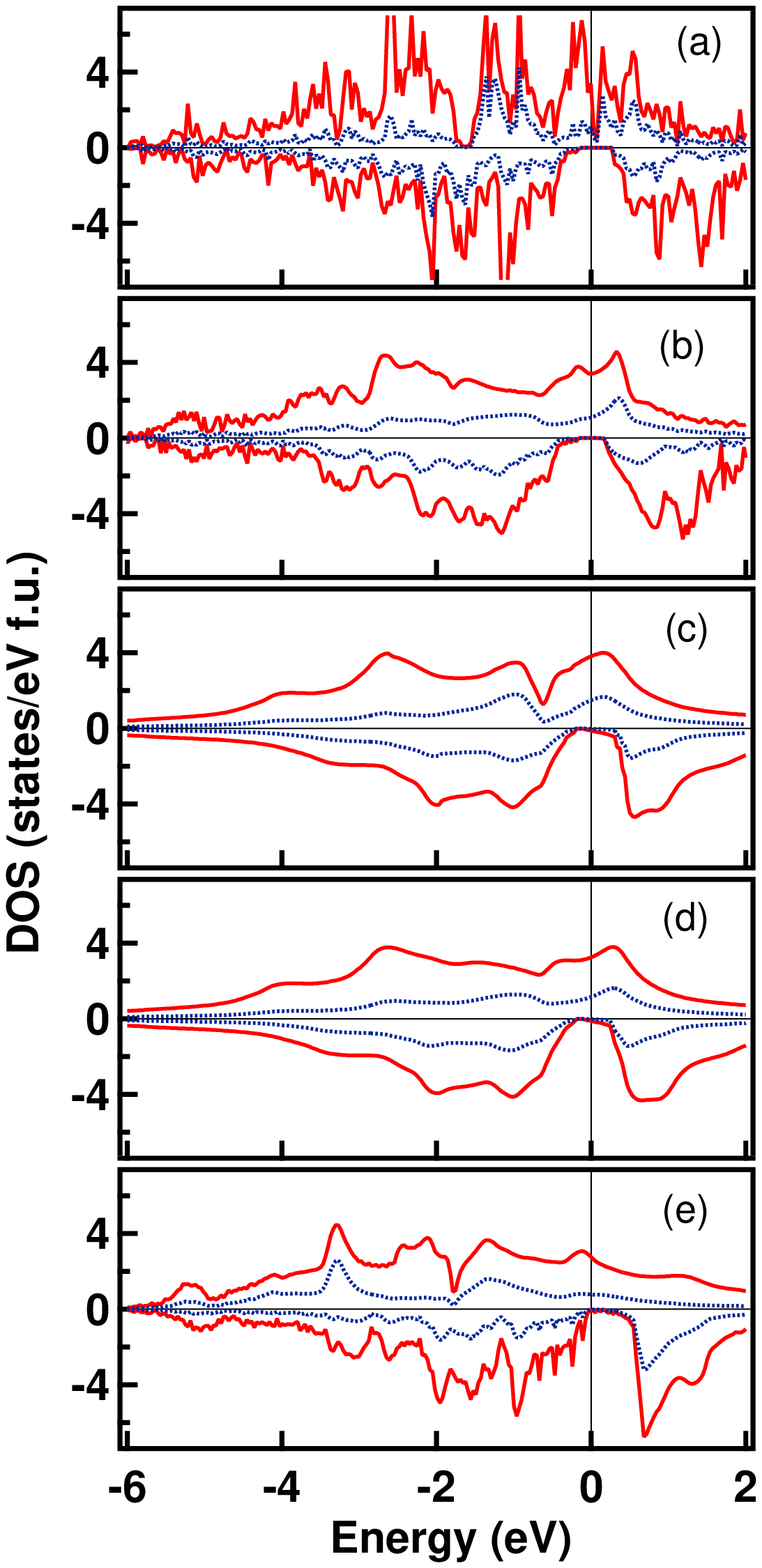}
\caption{
Calculated spin-resolved density of states (DOS) of CoFeCrAl with the various ordering phases: 
(a) full ordering $Y$, (b) full random swapping of Co and Fe, (c) full random swapping of Cr and Al, 
(d) full random swapping of Co and Fe as well as that of Cr and Al, and (e) full random swapping of Fe and Cr.
Dashed curves are the partial DOS for Fe.
}
\end{center}
\end{figure}

Figure 8 displays the results of ab-initio calculations of the element-specific magnetic moments, total magnetic moments, and formation energy for CoFeCrAl with various chemical orderings.
The theoretical data for the spin-resolved density of states (DOS) profiles for CoFeCrAl with various chemical orderings
are shown in Fig. 9.
The lattice parameter of CoFeCrAl was fixed to 0.575 nm in these calculations. 
The five cases of the chemical ordering and/or disordering considered here are as follows: (i) the full ordering [$Y$, Fig. 1(a)], (ii) the full random swapping of Co and Fe [$L2_1$, Fig. 1(c)], (iii) the full random swapping of Cr and Al [$L2_1$, Fig. 1(d)], (iv) the full random swapping of Co and Fe as well as that of Cr and Al [$B2$, Fig. 1(e)], and (v) the full random swapping of Fe and Cr [$XA$, Fig. 1(b)].
In all cases, the total magnetic moment $m$ is very close to 2.00 $\mu_B$/f.u. [Fig. 8(b)],
which is consistent with the predictions given by the Slater--Pauling--like rule observed in Heusler alloys with half-metallic gaps.
The half-metallic gap structures in the minority spin states survive in all cases, as seen in Fig. 9.
However, in some cases, finite DOS appear at around the Fermi level in the gap by the disorders [Figs. 9(c)-9(e)], 
meaning that the material is no longer a half-metal in a strict sense.

In case (i) (the ordered $Y$ structure),
the magnetic moment associated with the Fe atom, -0.703 $\mu_B$/atom, is antiparallel to that of the Co and Cr atoms (1.066 $\mu_B$/atom and 1.71 $\mu_B$/atom, respectively). Hence, there is an overall ferrimagnetic ground state, which is in good agreement with the literature.\cite{Choudhary2016}
In case (ii) (Cr-Al disorder),
the Fe atom, -0.227 $\mu_B$/atom, is antiferromagnetically coupled to both the Co and Cr atoms (0.889 $\mu_B$/atom and 
1.393 $\mu_B$/atom, respectively).
Additionally, we observe a similar kind of magnetic configuration in case (iii) (Co-Fe disorder), {\it i.e.}, 
the Fe atom has a magnetic moment alignment opposite to that of the Cr and Co atoms, 
and in case (iv), both of the above disorders (Co-Fe and Cr-Al) are simultaneously present in the system.
Thus, none of these cases reproduced the parallel arrangement of the magnetic moment of Fe and Co observed in XMCD, as summarized in Fig. 8(a).

In contrast, case (v) (disorder between Fe-Cr) qualitatively reproduced the abovementioned XMCD results.
The respective net moments of Fe and Co are 1.488 and 0.814 $\mu_B$/atom, respectively, and have a parallel configuration,
whereas Cr exhibits negligible net moment, as seen in Fig. 8(a).
The magnetic moments of Fe and Cr atoms at sites X' (Y) and Y (X) are 0.268 (2.708) $\mu_B$/atom and 1.318 (-1.682) $\mu_B$/atom, respectively. That is, Cr has two opposite magnetic moments at different sites that tend to cancel each other out. 
Here, the separation between the Cr at site X' and the Cr at site Y is around 0.249 nm, 
which is very much comparable to the separation of 0.248 nm in its bulk configuration.
This may be why the antiferromagnetic coupling between two nonequivalent Cr atoms as that of its bulk configuration.

When the CoFeCrAl is in the $Y$ phase, the SGS state was obtained, as described in a previous report [Fig. 9(a)].\cite{Xu2013}
In the case of Co-Fe disorder, the pseudo-gap in the majority spin band disappears, but half metallicity in the minority spin band is still observed [Fig. 9(b)]. 
However, similar to the other cases, Fe-Cr disorders destroy this half-metallicity, as mentioned above,
so that no large TMR effect is expected [Fig. 9(e)]. 
This is consistent with the observations in this study
if we suppose that our CoFeCrAl is similar to that with Fe-Cr disorders.
Furthermore, the CoFeCrAl with Fe-Cr disorders has a large magnetic moment of Fe at site Y that runs parallel to that of Co at site X.
This value of the magnetic moment for Fe is similar to that in Co$_2$FeAl Heusler alloys.
As seen in Fig. 9, 
the DOS peak for Fe is present at energies higher than the Fermi level when Fe is at site X or X' [Figs. 9(b)-9(d)],
except in the case of Fe-Cr ordering.
In contrast, the partial DOS of Fe in Fe-Cr disordered CoFeCrAl is more broad,
as compared with that for $Y$ and other cases, 
indicating that the energy band derived from the orbital of Fe is more similar to that for CoFe in Fe-Cr disordered CoFeCrAl.
Thus, the observation of CoFe/MgO-like coherent tunneling in this study
could be understood by considering the effect of the Fe-Cr disorder in terms of the partial DOS of Fe.
From the viewpoint of the formation energy, the $Y$-order state is most stable and the Fe-Cr disorder state is the most unstable. 
Note that all these calculations result in a ground state for the bulk,
whereas the experiments were conducted on films at room temperature.

Finally, it is appropriate to comment on the tunneling spin polarization for reference.
In many studies, Julliere's model is used to estimate the tunneling spin polarization, even in the coherent tunneling regime. This can be expressed as\cite{Julliere1975}
\begin{equation}
{\rm TMR \ ratio \ (\%)} = \frac{2 P_1 P_2}{1 - P_1 P_2} \times 100,
\end{equation}
where $P_1$ and $P_2$ are the tunneling spin polarizations for the respective magnetic electrodes.
If we assume $P$ = 0.85\cite{Parkin2004} or 0.69\cite{Marukame2006,Marukame2007} for CoFe in the coherent tunneling case observed in the MgO/CoFe system, for example,
then we obtain a $P$ value of 0.53 or 0.66, respectively, for $B2$-ordered CoFeCrAl from the highest TMR ratio at 10 K [165\% in this study, [Fig. 5(a)].
These values are similar to those obtained by PCAR, as mentioned in the Introduction.
However, they are low as compared with the values evaluated for Co$_2$-Heusler alloys with similar constituent elements, 
such as $P$ = 0.88 for Co$_2$Cr$_{0.6}$Fe$_{0.4}$Al.\cite{Marukame2007}
Future research will investigate the TMR effect and spin polarization of CoFeCrAl with much higher chemical orderings of $L2_1$ or $Y$.

\section{Summary}
Fully epitaxial (001)-oriented MTJs with CoFeCrAl electrode film were grown on a Cr buffer.
The CoFeCrAl films had atomically flat surfaces and $B2$ chemical ordering, as confirmed by XRD and TEM measurements.
$M_{\rm s}$ = 380 emu/cm$^3$ was observed, corresponding to the value given by the Slater--Pauling-like rule.
The maximum TMR ratios were 87 and 165\% at 300 and 10 K, respectively.
The MTJs had MgO-interfaces with fewer dislocations, as observed by cross-sectional TEM measurements.
Both magnon-induced inelastic electron tunneling and coherent electron tunneling were suggested by the temperature- and bias-voltage-dependence measurements of the transport properties.
The ferromagnetic arrangement of the Co and Fe magnetic moments for the CoFeCrAl film was confirmed by XMCD measurements,
contrary to the ferrimagnetic arrangement predicted in the $Y$-ordered state possessing SGS characteristics.
Ab-initio calculations taking account the Cr-Fe swap disorder qualitatively explained these XMCD results.
We also discussed the effect of the Cr-Fe swap disorder on the electronic states, which allow coherent electron tunneling,
in terms of the partial DOS for Fe atoms.

\section*{Acknowledgements}
T.T and S.M. would like to thank Y. Kondo for his technical support. 
This work was partially supported by JST CREST (No. JPMJCR17J5), JSPS Core-to-Core Program, and KAKENHI (No. 17F17063).

\end{document}